\documentclass[%
twocolumn,
showpacs,
 amsmath,amssymb,
 aps,
prl,
floatfix,
]{revtex4-1}

\usepackage{graphicx, color}
\usepackage{dcolumn}
\usepackage{bm}
\usepackage{ulem}
\usepackage{xspace}
\usepackage[utf8]{inputenc}  
\usepackage[T1]{fontenc}

\newcolumntype{d}{D{.}{.}{-1}}    
\newcolumntype{b}{D{(}{\ (}{-1}}  

\newcommand{\eref}[1]{(\ref{#1})}

\newcommand{\Tref}[1]{Table~\ref{#1}}
\newcommand{\Fig}[1]{Fig.~\ref{#1}}

\newcommand{\Ca}[1]{\ensuremath{^{#1}}Ca\ensuremath{^{+}}}
\newcommand{\Mg}[1]{\ensuremath{^{#1}}Mg\ensuremath{^{+}}}
\newcommand{\mass}[1]{\ensuremath{m_\text{#1}}}
\newcommand{\unit}[2]{\ensuremath{#1}\hspace{2pt}{#2}}

\newcommand{\kms}{\ensuremath{k_\textrm{MS}}}
\newcommand{\fs}{\ensuremath{F}}
\newcommand{\drsq}{\ensuremath{\delta\!\left< r^2 \right>}}

\newcommand{\GHzamu}{GHz$\cdot$amu}
\newcommand{\MHzfmsq}{\textrm{MHz$\cdot$fm$^{-2}$}}
\newcommand{\fmsq}{\ensuremath{\textrm{fm}^{2}}}

\def\fm#1{\ifmmode #1 \else $#1$\fi}
\def\ket#1{{%
  \ifmmode |\,#1\,\rangle \else $|\,#1\,\rangle$\fi}}
\def\bra#1{{%
  \ifmmode \langle\,#1\,| \else $\langle\,#1\,|$\fi}}
\def\braket#1#2{{%
  \ifmmode \langle\,#1\,|\,#2\,\rangle \else $\langle\,#1\,|\,#2\,\rangle$\fi}}
\def\expect#1{{%
  \ifmmode \langle\,#1\,\rangle \else $\langle\,#1\,\rangle$\fi}}

\def\dsoh{\fm{{}^2\mathrm{S}_{1/2}}\xspace}
\def\dpoh{\fm{{}^2\mathrm{P}_{1/2}}\xspace}

\def\ddth{\fm{{}^2\mathrm{D}_{3/2}}\xspace}

\def\dsohdpoh{\dsoh\fm{\rightarrow}\enskip\dpoh\xspace}
\def\ddthdpoh{\ddth\fm{\rightarrow}\enskip\dpoh\xspace}

\begin{document}
\title{Precision isotope shift measurements in Ca$^+$ using highly sensitive detection schemes}




\author{Florian Gebert$^1$}
\author{Yong Wan$^1$}%
\author{Fabian Wolf$^1$}%
\author{Christopher N. Angstmann$^2$}
\author{Julian C. Berengut$^3$}
\author{Piet O. Schmidt$^1$}%
\affiliation{%
 $^1$QUEST Institute, Physikalisch-Technische Bundesanstalt, 38116 Braunschweig, Germany\\
   $^2$School of Mathematics and Statistics, University of New South Wales, Sydney NSW 2052, Australia\\
 $^3$School of Physics, University of New South Wales, Sydney, NSW 2052, Australia
}%

%

\date{\today}

\begin{abstract}
We demonstrate an efficient high-precision optical spectroscopy technique for single trapped ions with non-closed transitions. In a double-shelving technique, the absorption of a single photon is first amplified to several phonons of a normal motional mode shared with a co-trapped cooling ion of a different species, before being further amplified to thousands of fluorescence photons emitted by the cooling ion using the standard electron shelving technique. We employ this extension of the photon recoil spectroscopy technique to perform 
the first high precision absolute frequency measurement of the \ddthdpoh transition in \Ca{40}, resulting in a transition frequency of $f=346\, 000\, 234\, 867(96)$~kHz. Furthermore, we determine the isotope shift of this transition and the \dsohdpoh transition for \Ca{42}, \Ca{44} and \Ca{48} ions relative to \Ca{40} with an accuracy below 100\,kHz. Improved field and mass shift constants of these transitions as well as changes in mean square nuclear charge radii are extracted from this high resolution data.


\end{abstract}


\maketitle

\begin{figure}[tb]
	\centering	\includegraphics[width=0.50\textwidth]{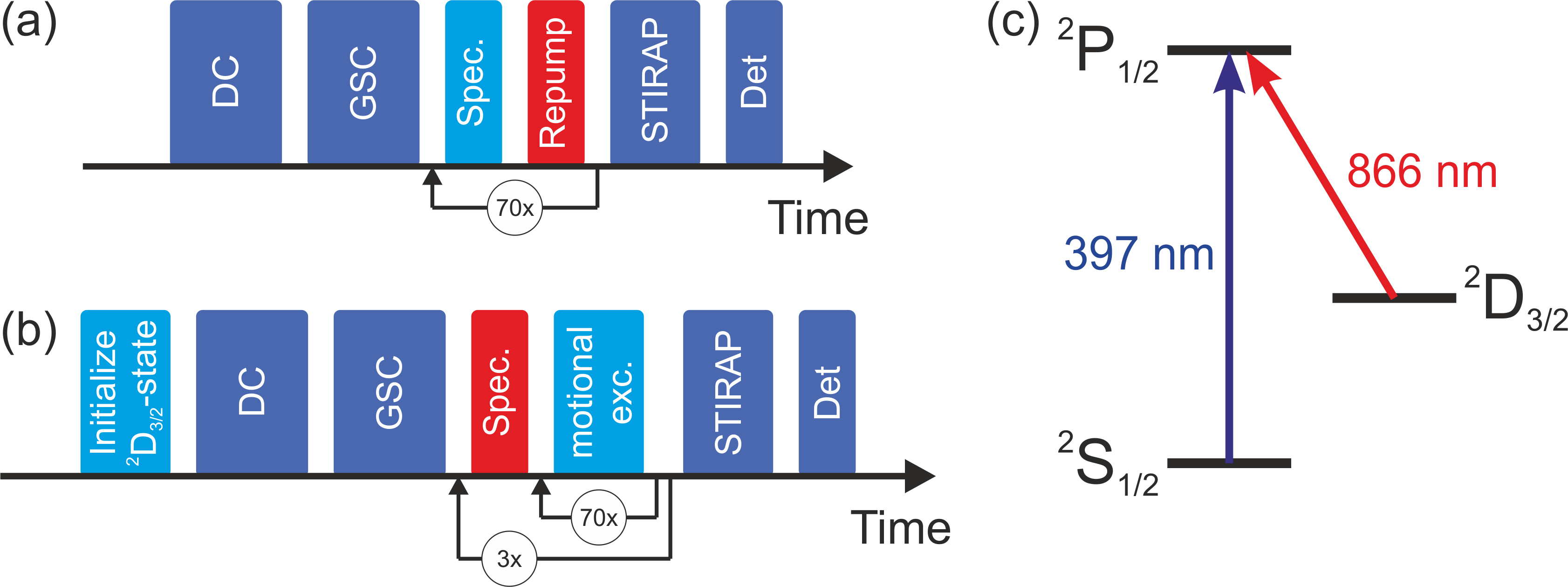}
	\caption{Experimental sequence for the spectroscopy on (a) the \dsohdpoh transition and (b) the \ddthdpoh transition. A detailed description is given in the text. (c) Simplified level scheme of \Ca{} ions}
	\label{fig:experimental_sequence}
\end{figure}
\begin{figure}[tb]
	\centering
		\includegraphics[width=0.50\textwidth]{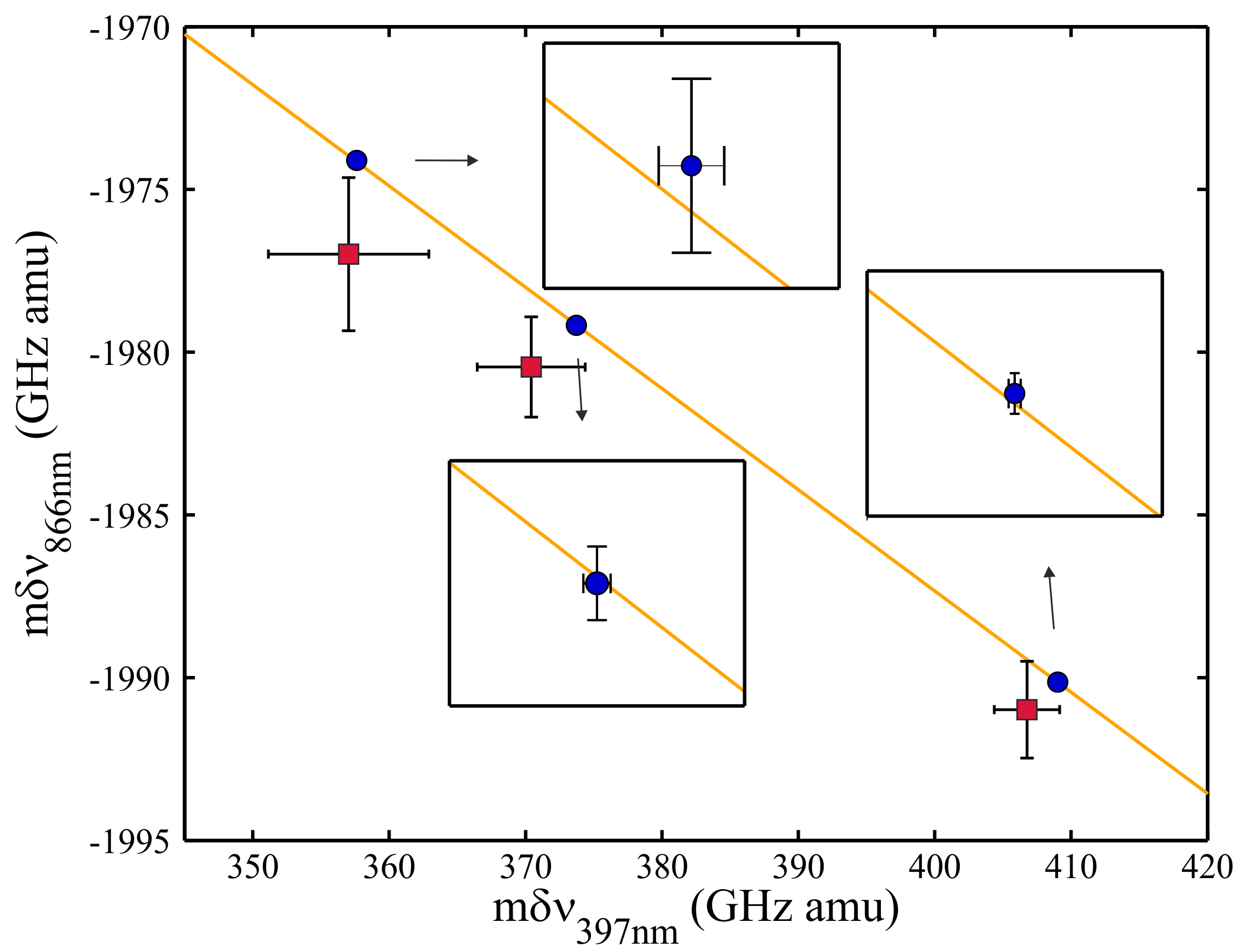}
	\caption{Two-dimensional King plot showing modified isotope shift of the 866 and 397 lines. Red squares: previous experimental data from \cite{ma_artensson-pendrill_isotope_1992} and \cite{nortershauser_isotope_1998}; blue circles: this work. The insets show the relevant ranges enlarged by a factor of approximately 30 to illustrate the quality of the fit.}
	\label{fig:King_plot}
\end{figure}

Single or few laser-cooled ions trapped in Paul traps allow spectroscopy with high resolution and accuracy due to the near perfect realization of an unperturbed transition frequency as demonstrated in ion-based optical clocks \cite{ludlow_optical_2014}. For long-lived excited states the electron shelving technique provides near unity detection efficiency \cite{dehmelt_shelved_1975}. Advanced laser fluorescence techniques are well-suited for closed or nearly-closed broad, dipole-allowed transitions \cite{gardner_precision_2014, herrmann_frequency_2009}. However, they provide insufficient signal to efficiently study non-closed transitions where typically only one photon can be absorbed before the ion decays to a state which is not addressed by the spectroscopy laser. As a result, the fluorescence rate is low even when pump-repump techniques are employed, resulting in long averaging times \cite{pruttivarasin_direct_2014, wubbena_controlling_2014}. Recently, new spectroscopy techniques with a high sensitivity to the absorption of single or few photons have been developed employing the signal from photon recoil \cite{hempel_entanglement-enhanced_2013, wan_precision_2014} or decoherence of an electronic superposition state \cite{clos_decoherence-assisted_2014}. An evaluation of the dominant effects shifting the resonance from its unperturbed frequency provides accuracies below 100~kHz \cite{wan_precision_2014, herrmann_frequency_2009}. Up to now, these absolute frequency measurements have been demonstrated on nearly-closed transitions.\\
Here, we demonstrate an extension of the photon recoil spectroscopy (PRS) technique \cite{wan_precision_2014} to perform an absolute frequency measurement of the non-closed \ddthdpoh transition in \Ca{40} with an accuracy of below 100~kHz. Using this extension together with the original PRS scheme, we performed isotope shift measurements of the \ddthdpoh and \dsohdpoh transitions for the most abundant even \Ca{} isotopes with an uncertainty of below 100~kHz. These measurements improve the accuracy of the isotope shifts of the \Ca{} ions by up to two orders of magnitude compared to previous measurements \citep{ma_artensson-pendrill_isotope_1992, nortershauser_isotope_1998}.\\
From a multi-dimensional King plot analysis \cite{palmer_laser_1984, angeli_table_2013}, seeded by the changes in mean square nuclear charge radii obtained from muonic atom spectroscopy and electron scattering, we derive improved values for the field and mass shift constants as well as the change in the mean square nuclear charge radius. The high accuracy of these measurements serves as a benchmark for \textit{ab initio} atomic structure calculations and enables improved calibration for collinear laser spectroscopy \cite{silverans_fast_1985, nortershauser_isotope_1998}, where typically uncertainties of a few MHz are achieved. These measurements are relevant for the investigation of nuclear properties of unstable calcium isotopes \cite{garcia_ruiz_ground-state_2015, gorges_to_be_submitted_2015}.\\
Calcium is of particular interest since its isotopes include two doubly magic nuclei at $A=40$ and $48$. In the shell model, the additional 8 neutrons form a complete $1f_{7/2}$ shell, however there is little change in the mean-square proton radius (charge radius). Understanding these peculiar changes in the calcium charge radius is a challenge for nuclear theory.\\
We use the experimental setup for PRS with \Mg{25} as the cooling ion species described in detail in \cite{hemmerling_single_2011, wan_precision_2014, wan_efficient_2015}. In brief, we prepare two-ion crystals in a linear Paul trap consisting of \Mg{25} and a selected \Ca{} isotope as the spectroscopy ion. Isotope-selective two-photon ionization is used to load the different \Ca{} isotopes: the $^{1}$S$_{1}$\xspace\fm{\rightarrow}\enskip$^{1}$P$_{1}$\xspace transition of the neutral species of the target isotope is selectively excited using a narrow-linewidth laser \citep{lucas_isotope-selective_2004}, followed by ionization with a second laser. After loading, the isotopes are identified by the motional normal mode frequencies $\omega_{\mathrm{i,o}}$ (i: in phase, o: out of phase) of the two-ion crystal in the axial trapping potential, given by~\citep{drewsen_nondestructive_2004, wuebbena12pra},
\begin{equation}
\omega_{\mathrm{i,o}}^{2}=\omega_{\mathrm{z}}^{2}\cdot\left(1+\mu\mp\sqrt{1-\mu+\mu^2}\right),	
\label{eqn:trap_freq}
\end{equation}
where $\omega_z$ is the axial normal mode frequency of a single \Mg{25}, $\mu=\mass{Mg}/\mass{Ca}$, and \mass{Mg} and \mass{Ca} are the masses of the two ions in the trap. Following the technique described in \citep{drewsen_nondestructive_2004}, the normal mode frequencies were measured and fits of the in-phase normal mode frequencies lead to \unit{1.922(1)}{MHz}, \unit{1.889(1)}{MHz}, \unit{1.857(1)}{MHz} and \unit{1.795(1)}{MHz} for a two-ion crystal containing \Ca{40}, \Ca{42}, \Ca{44} and \Ca{48}, respectively. The measured frequencies are in agreement with the frequencies calculated from (\ref{eqn:trap_freq}) for the measured axial normal mode frequency of a single \Mg{25} of \unit{2.221(1)}{MHz}.\\
Isotope shift measurements for the \dsohdpoh transition of the different \Ca{} isotopes were performed using the original PRS scheme \citep{wan_precision_2014}. The experimental sequence of the technique is displayed in \Fig{fig:experimental_sequence}a). After Doppler cooling, we further cool the axial modes of the two-ion crystal to the motional ground state via sideband cooling on \Mg{25} \citep{wan_efficient_2015}. 
Starting from the motional ground state of one of the axial normal modes of the two-ion crystal, the spectroscopy laser with a wavelength of \unit{397}{nm} is applied in \unit{125}{ns} short pulses. The laser pulse repetition is matched to the motion of the two-ion crystal, thus enhancing the sensitivity through resonant driving and enabling the detection of around 10 absorbed photons with a signal-to-noise ratio of 1 \citep{wan_precision_2014}. A repump laser at \unit{866}{nm} resonant with the \ddthdpoh transition is applied interleaved with the spectroscopy laser pulses to prevent populating the metastable \ddth-state. These spectroscopy/repump cycles are repeated 70 times, obtaining the maximum motional excitation detectable without saturating the signal. The residual ground state population is measured using a red sideband stimulated Raman adiabatic passage (STIRAP) pulse sequence coupling the two ground state hyperfine levels in \Mg{25} \cite{gebert_population_2015}. This maps the motional excitation efficiently into electronic excitation, detectable via discrimination of the \Mg{25} hyperfine states using electron shelving with the $\pi$-detection technique \cite{hemmerling_novel_2012}.\\
We extend the original PRS scheme to measure the absolute frequency and the isotopic shifts of the open \ddthdpoh transition in \Ca{} as shown in \Fig{fig:experimental_sequence}(b). The sequence starts by initializing the \Ca{} ion in the \ddth state via optical pumping on the \dsohdpoh transition and subsequent sideband cooling on \Mg{25}. Absorption of a photon on the \ddthdpoh spectroscopy transition in \Ca{} results in a decay to the \dsoh ground state with 94~\% probability \cite{ramm_precision_2013}, where the ion is lost from the normal spectroscopy cycle. Starting from the ground state of motion of one of the ion's normal modes, absorption of photons from a pulsed laser tuned to resonance with the \dsohdpoh transition in \Ca{} efficiently excites this motional mode analogously to the original PRS technique \cite{wan_precision_2014}. This corresponds to a first shelving step, since the excitation of phonons is conditional upon the absorption of a photon on the spectroscopy transition. Due to the non-vanishing probability of the ion to decay to the \ddth state the ion is effectively reinitialized in this state after 70 excitation pulses. Repeating the spectroscopy/motional excitation cycle three times further enhance the motion of the two-ion crystal to obtain the maximum motional excitation without reaching saturation. The motional excitation is mapped onto the \Mg{25} ion and detected as before.\\
Commercially available external cavity diode lasers are used to excite the transitions of the different \Ca{} isotopes, where the laser coupling the \dsoh and the \dpoh state is frequency doubled to \unit{397}{nm}. Both laser beams used in the two spectroscopy techniques are aligned spanning a $45^{\circ}$ angle with an external magnetic field required for efficient cooling and detection. The polarization is aligned perpendicular to the magnetic field to couple all magnetic substates of the involved transitions. The frequencies of both lasers are stabilized in the infrared to the frequency doubled output of a fiber-based optical frequency comb using an electronic feedback loop. A relative frequency stability of better than $10^{-12}$ is achieved with respect to the comb for averaging times of a second. The comb is stabilized to the \unit{10}{MHz} reference signal obtained from a hydrogen maser, which is referenced to the SI second at the German Metrology Institute (PTB).\\
The lasers are steered to the center frequencies of the spectroscopy transitions using acousto-optical modulators in double-pass configuration. 
We apply the two-point sampling technique \citep{wan_precision_2014}, where the population difference at half maximum below and above the resonance line is measured and the difference is used to correct the center of the two probing frequencies for the next scan. The population difference in combination with the experimentally determined slope at these positions is used in post evaluation to derive the center frequency of the atomic transition.\\
Following \cite{wan_precision_2014} we evaluated the dominant effects which could result in a shift of the observed transition frequency from its unperturbed value. External magnetic fields in combination with polarization asymmetries can cause an asymmetric line shape shifting the center frequency of the transition. Furthermore, the electrical field from the spectroscopy laser light may shift the frequency by means of the ac stark shift, even though small intensities on the order of a few $\%$ of the saturation intensity were employed for probing. Therefore, we measured the transition frequency for different magnetic fields and different spectroscopy laser intensities and extracted the corresponding shifts $\Delta f_\text{Zeeman}=$ \unit{54\pm44}{kHz} and $\Delta f_\text{Stark}=$ \unit{-70\pm83}{kHz} for the experimentally used parameters. The given uncertainties correspond to the \unit{68.3}{\%} prediction bound of the fit to the measured data and are limited by statistics. Due to the first shelving step in the newly developed technique, the dominant shift in the PRS technique, the so-called lineshape shift arising from probe-frequency-dependent Doppler cooling and heating effects during motional excitation, is expected to be small. This was confirmed using density matrix simulations, where the shift was determined to be well below \unit{1}{kHz}. The absolute frequency of the \ddthdpoh transition was derived as the weighted average of $6$ frequency measurements and is given by $f=346\,000\,234\,867(96)$~kHz, where \unit{94}{kHz} of the uncertainty results from systematic and \unit{21}{kHz} from statistical uncertainties.\\
We used the original and extended PRS technique to perform isotope shift measurements for the \dsohdpoh and \ddthdpoh transitions, respectively. The measurements of each isotope and the reference \Ca{40} were performed interleaved. We first measured the \Ca{40} transition frequency. Then, we loaded a new two-ion crystal containing the investigated isotope of \Ca{} and measured its transition frequency. After this, we repeated the measurement of the reference ion \Ca{40} to ensure the absence of frequency drifts.\\
\begin{figure}[tb]
	\centering
		\includegraphics[width=0.50\textwidth]{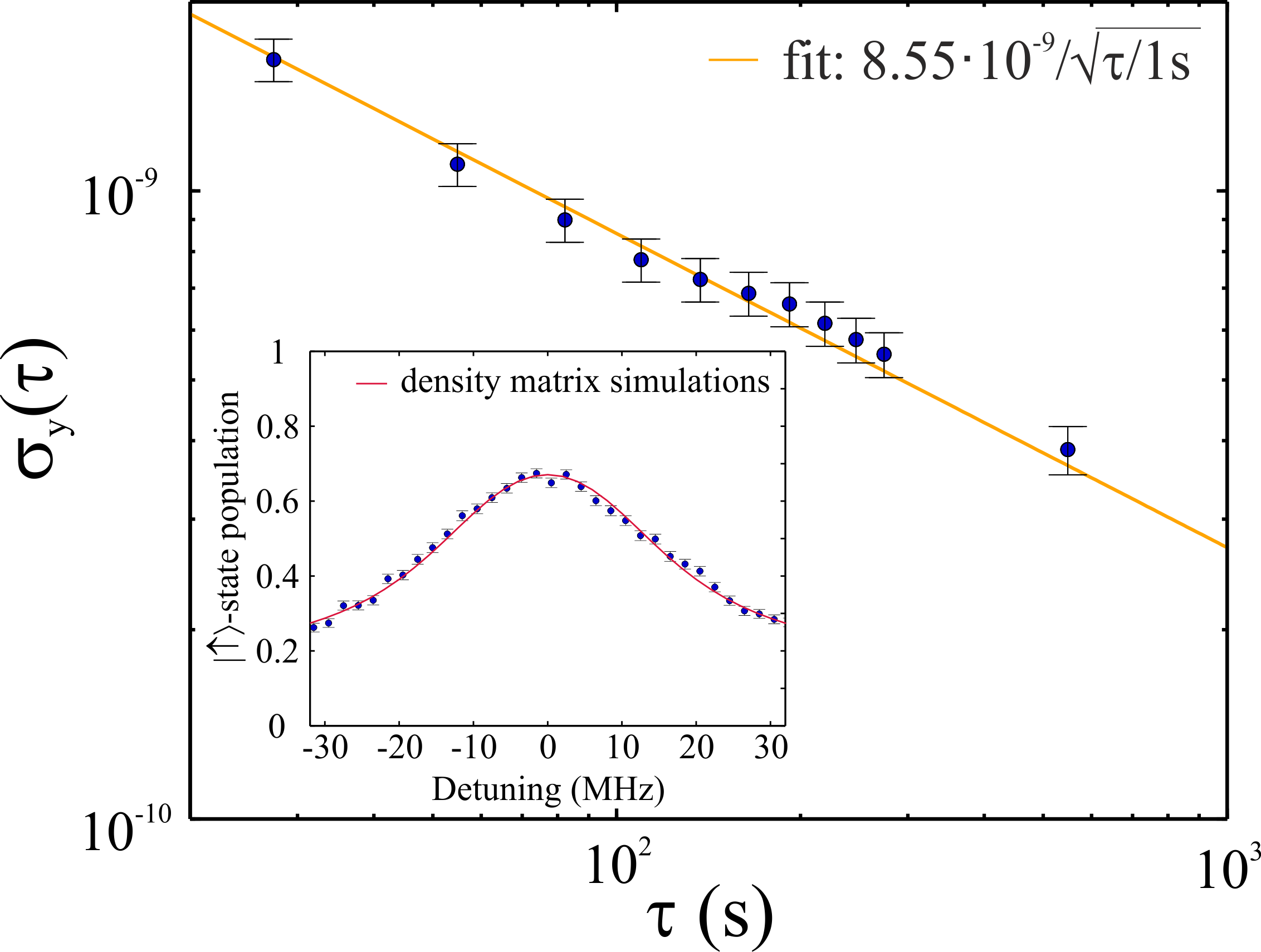}
	\caption{Overlapped Allan deviation $\sigma_y(\tau)$ of the frequency measurement of the \ddthdpoh transition of \Ca{40}. The inset shows a resonance scan of the transition, where the red line corresponds to density matrix simulations adjusted in amplitude and offset to account for experimental imperfections.}
	\label{fig:allan_deviation_overlap_Ca40}
\end{figure}
We investigated the statistical properties of all frequency measurements by plotting the overlapped Allan deviation $\sigma_y(\tau)$. No indication of drifts or other frequency fluctuations were observed for the performed measurements. As an example, the overlapped Allan deviation for the measurement of the \ddthdpoh transition of \Ca{40} is shown in \Fig{fig:allan_deviation_overlap_Ca40}. After averaging, the measured frequencies were used to derive the isotope shift for each measurement. For the \dsohdpoh transition the measured isotope shifts were corrected by the change of the probe frequency-dependent lineshape shift, resulting from the different mass of the isotope and the corresponding change of the normal mode frequency of the two-ion crystal. Using the simple model derived in \citep{wan_precision_2014}, the calculated correction is smaller than \unit{2}{kHz} per mass unit corresponding to a maximum correction of \unit{10}{kHz}. All other systematic effects are common for the measurements performed on the same day day and are therefore removed. The final values $\delta\nu^{A,40}$ with the corresponding accuracies are displayed in \Tref{tab:isotope shifts}.\\
\begin{table}[b]
\caption{Measured isotope shifts of the two transitions referenced to \Ca{40}.}
\label{tab:isotope shifts}
\begin{ruledtabular}
\centering
\begin{tabular}{cbb}
A & \multicolumn{1}{c}{$\delta\nu^{A,40}_{397}$ (MHz)} & \multicolumn{1}{c}{$\delta\nu^{A,40}_{866}$ (MHz)} \\
\hline
42 &  425.706(94)  & -2349.974(99) \\
44 &  849.534(74) & -4498.883(80) \\
48 & 1705.389(60) & -8297.769(81) \\
\end{tabular}
\end{ruledtabular}
\end{table}
Isotope shifts in atomic transition frequencies originate from two effects: a change in the size of the nucleus and thus its interaction with the electrons (field shift), and a change in the recoil of the nucleus (mass shift). The mass shift is more important for light elements, while for heavy elements the field shift dominates \citep{breit_theory_1958}. The difference in the transition frequency, $\delta\nu^{A,A'}$, between isotopes with mass~$m_A$ and~$m_{A'}$ can be expressed as
\begin{equation}
\label{eq:IS}
\delta \nu^{A,A'} = \kms \left(\frac{1}{m_{A}} - \frac{1}{m_{A'}}\right) + \fs \drsq^{A, A'} \,,
\end{equation}
where \kms\ is the mass shift constant, \fs\ is the field shift constant, and $\delta\!\left< r^2 \right>$ is the change in the mean-square nuclear charge radius. The latter is a common parameter when comparing two different transitions and can be eliminated in a King plot analysis as shown in \Fig{fig:King_plot} for the two transitions considered here. Each axis shows the modified isotope shift $m\delta\nu^{A,A'}=\delta \nu^{A,A'}g^{A,A'}$, where $g^{A,A'} = (1/m_{A} - 1/m_{A'})^{-1}$, for one of the two transitions. A straight line fit to the three data points provides linear combinations of the field and mass shift constants for the two transitions. An important result from this fit is that there is no evidence for a deviation from a straight line, confirming that \eref{eq:IS} is a good parametrization of the isotope shift even at the high experimental accuracy of the measurements presented here.\\
A comparison of the high resolution results with previous experimental data based on collinear laser spectroscopy \cite{ma_artensson-pendrill_isotope_1992, nortershauser_isotope_1998} shows systematic deviations, which can be used to calibrate experimental parameters of this technique. 
Following Ref.~\cite{palmer_laser_1984} we performed a three-dimensional King plot analysis to extract the fitting parameters \kms\ and \fs\ for the two transitions. Two dimensions are those shown in \Fig{fig:King_plot}. In the third dimension we plot the modified change in mean-square nuclear charge radius $\drsq^{A,A'}g^{A,A'}$, using the previous values of \drsq\ from~\cite{wohlfahrt_muonic_1978}, which are based on muonic atom spectroscopy and electron scattering. The three-dimensional King plot constrains the mass and field-shift constants, and under the assumption that \eref{eq:IS} is correct (i.e. the three data points are connected by a straight line) can also be used to extract improved values of \drsq. To find the parameter estimates and their uncertainties an acceptance/rejection Monte Carlo method was used to generate samples consistent with the measured values and associated uncertainties. The measurement distributions were assumed to be independent uncorrelated normals. The likelihoods of three randomly generated points, constrained to be collinear, were used as the acceptance criterion in the algorithm. The extracted parameters are shown in \Tref{tab:parameters}.\\
\begin{table}[tb]
\caption{Parameters of three-dimensional King plot seeded with values of $\drsq^{A,40}$ taken from \cite{wohlfahrt_muonic_1978}. The units for the field $F_i$ and mass $k_i$ shift constants and the changes in mean square nuclear charge radii $\drsq^{j,40}$ are \MHzfmsq, \GHzamu \hspace{2pt} and \fmsq, respectively. For comparison the second column for the previous data shows results for the analysis using isotope shift data taken from \cite{ma_artensson-pendrill_isotope_1992} and \cite{nortershauser_isotope_1998} analyzed with the methods used in this work. \label{tab:parameters}}
\begin{ruledtabular}
\begin{tabular}{lcccc}
Param. & \multicolumn{2}{c}{Previous} & \multicolumn{1}{c}{This work} & \multicolumn{1}{c}{Theory} \\
\hline
$F_{397}$  & $-283(6)$\footnotemark[1] &$-281(34)$& $-281.8(7.0)$ & $-285(3)$\footnotemark[1] \\
                                       &&&& $-287$\footnotemark[2] \\
$k_{397}$  & $405.1(3.8)$\footnotemark[1] & $406.4(2.8)$ & $408.73(40)$ & $359$\footnotemark[2] \\
                                       &&&& $427$\footnotemark[4] \\
$F_{866}$  & $79(4)$\footnotemark[3] & $80(13)$ & $87.7(2.2)$ & $88$\footnotemark[1] \\
                                       &&&& $92$\footnotemark[2] \\
$k_{866}$  & $-1989.8(4)$\footnotemark[3] & $-1990.9(1.4)$ & $-1990.05(13)$ & $-2207$\footnotemark[2] \\
                                       &&&& $-2185$\footnotemark[4] \\
$\drsq^{42,40}$  & $0.210(7)$ & $0.210(7)$ & $0.2160(49)$ \\
$\drsq^{44,40}$  & $0.290(9)$ & $0.290(9)$ & $0.2824(65)$ \\
$\drsq^{48,40}$  & $-0.005(6)$ & $-0.005(6)$ & $-0.0045(60)$ \\
\end{tabular}
\end{ruledtabular}
\footnotetext[1]{M{\aa}rtensson-Pendrill \textit{et al.}, 1992 \cite{ma_artensson-pendrill_isotope_1992}.}
\footnotetext[2]{Safronova and Johnson, 2001 \cite{safronova_third-order_2001}.}
\footnotetext[3]{N\"ortersh\"auser \textit{et al.}, 1998 \cite{nortershauser_isotope_1998}.}
\footnotetext[4]{This work, based on the methods in \cite{berengut_isotope-shift_2003}.}
\end{table}
The extracted field-shift and mass-shift constants pose a strong challenge for many-body atomic theory (fourth column of \Tref{tab:parameters}), where the mass-shift in particular has proven very difficult to calculate even in the ``easy'' case of single-valence-electron ions \cite{safronova_third-order_2001,berengut_isotope-shift_2003}. A comparison to the experimental field and mass shift constants given in \cite{ma_artensson-pendrill_isotope_1992, nortershauser_isotope_1998} proves difficult since the derived uncertainties depend strongly on the analysis technique and input parameters for \drsq. Evaluating the field and mass shift constant from isotope shifts given in \cite{ma_artensson-pendrill_isotope_1992, nortershauser_isotope_1998} using the analysis and input parameters employed here, indicates an order of magnitude improvement in these values based on the high resolution data presented here. Furthermore, the analysis leads to improved values of changes in mean-square nuclear charge radii which are in good agreement with the values derived from isotope shift measurements in neutral Ca \cite{palmer_laser_1984}.\\
In conclusion, we presented the first high precision absolute frequency measurement of the \ddthdpoh transition in \Ca{} using an extension of photon recoil spectroscopy. We measured isotope shifts of this and the \dsohdpoh transition of the most abundant stable, even isotopes in \Ca{} using this highly sensitive technique and improved the accuracy of the shifts by up to two orders of magnitude. From these measurements, improved values of the field and mass shift constants as well as changes in mean square nuclear charge radii are extracted. These results may serve to validate calculations and determine optimal models used in isotope shift calculations. Furthermore, the precise knowledge of the isotopic shifts can be used to improve the calibration of collinear laser spectroscopy experiments \cite{garcia_ruiz_ground-state_2015, gorges_to_be_submitted_2015} and improve the analysis of quasar absorption spectra aimed at searches for variations of fundamental constants \cite{murphy_laboratory_2014}. This work demonstrates that photon recoil spectroscopy is a versatile technique, suitable for a wide range of high precision spectroscopic applications.\\
\begin{acknowledgements}
We acknowledge the support of DFG through QUEST and grant SCHM2678/3-1. This work was financially supported by the State of Lower-Saxony, Hannover, Germany.  Y. W. acknowledges support from IGSM. We thank R. Blatt for generous loan of equipment, W. N\"ortersh\"auser and C. Elster for stimulating discussions concerning the data evaluation, and J. B. W\"ubbena, I. D. Leroux and N. Scharnhorst for technical assistance with the absolute frequency measurement.
\end{acknowledgements}
\bibliography{references}

\end{document}